\documentclass[twocolumn]{aastex63}

\received{2020 January 23}
\revised{2020 March 12}
\accepted{2020 March 16}

\submitjournal{ApJL}

\shorttitle{Time Variations of Sgr A* at 230 GHz}
\shortauthors{Iwata et al.}

\begin{document}

\title {Time Variations in the Flux Density of Sgr A* at 230 GHz Detected with ALMA}

\correspondingauthor{Yuhei Iwata}
\email{iwa-astro4338@z3.keio.jp}

\author[0000-0002-9255-4742]{Yuhei Iwata}
\affiliation{School of Fundamental Science and Technology, Graduate School of Science and
Technology, Keio University, 3-14-1 Hiyoshi, Kohoku-ku, Yokohama, Kanagawa 223-8522, Japan}
\author[0000-0002-5566-0634]{Tomoharu Oka}
\affiliation{School of Fundamental Science and Technology, Graduate School of Science and
Technology, Keio University, 3-14-1 Hiyoshi, Kohoku-ku, Yokohama, Kanagawa 223-8522, Japan}
\affiliation{Department of Physics, Faculty of Science and Technology, Keio University, 3-14-1
Hiyoshi, Kohoku-ku, Yokohama, Kanagawa 223-8522, Japan}
\author[0000-0001-8185-8954]{Masato Tsuboi}
\affiliation{Institute of Space and Astronautical Science(ISAS), Japan Aerospace Exploration Agency, 3-1-1 Yoshinodai, Chuo-ku, Sagamihara, Kanagawa 252-5210, Japan}
\affiliation{Department of Astronomy, The University of Tokyo, Bunkyo, Tokyo 113-0033, Japan}
\author[0000-0002-6272-507X]{Makoto Miyoshi}
\affiliation{Jasmine Project, National Astronomical Observatory of Japan, 2-21-1 Osawa, Mitaka, Tokyo 181-8588, Japan}
\affiliation{Department of Astronomical Science, School of Physical Sciences, The Graduate University for Advanced Studies (SOKENDAI), 2-21-1 Osawa, Mitaka, Tokyo 181-8588, Japan}

\author[0000-0001-8147-6817]{Shunya Takekawa}
\affiliation{Nobeyama Radio Observatory, National Astronomical Observatory of Japan, 462-2 Nobeyama, Minamimaki, Minamisaku-gun, Nagano 384-1305, Japan}

\begin{abstract} 
A radio source at the Galactic center Sgr A* is a prime supermassive black hole candidate and therefore key to developing our understanding of them. Time variations in the 230 GHz band flux of Sgr A* have been found with the Atacama Large Millimeter/submillimeter Array (ALMA) Cycle 5 observations. Measuring the flux density of Sgr A* in 1 min snapshots at 217.5, 219.5, and 234.0 GHz, we obtained light curves for ten 70 min periods. The light curves show variations at a few tens of minutes and hourly scales. The shorter timescale is similar to the orbital period of the innermost stable circular orbit around a $4\times 10^{6}$ $M_{\sun}$ black hole, suggesting that the variation originates from the immediate vicinity of Sgr A*. We also detected no time lag between 217.5 and 234.0 GHz and a dependence of the spectral index on the flux density. 

\end{abstract}


\section{Introduction} \label{sec:intro}

Sgr A* is a bright and compact radio source at the Galactic center and is one of the most convincing candidates for supermassive black holes. The mass of Sgr A* has been measured to be $4\times10^6\>M_{\sun}$ through analysis of the orbital parameters of the fast-moving star S2 \citep[e.g.,][]{Gravity18a,Do19a}. The radio band of Sgr A* reveals a positive spectral index $\alpha$ (where $F_{\nu} \propto \nu^{\alpha}$) that peaks around the ``submillimeter bump'' \citep{Falcke98}, at which point the flux rapidly drops to the infrared regime. Its spectral energy distribution (SED), low bolometric to Eddington luminosity ratio \citep[$L/L_{\rm EDD} \sim 10^{-9}$;][]{Genzel10}, and low accretion rate \citep[$\la 10^{-7}\>M_{\sun}\>{\rm yr^{-1}}$;][]{Marrone06} are often interpreted in the context of radiatively inefficient accretion flow models \citep{Yuan03}.

Sgr A* shows variability and flaring activity in the radio \citep[e.g.,][]{Miyazaki04,Macquart06,Yusef-Zadeh11,Brinkerink15}, near-infrared (NIR) \citep[e.g.,][]{Genzel03,Do19b}, and X-ray bands \citep[e.g.,][]{Baganoff01,Haggard19}. Its variation timescales have been reported to be a few tens of minutes, which may be related to quasi-periodic oscillations (QPOs) occurring near the innermost stable circular orbit (ISCO) of Sgr A* \citep{Genzel03,Aschenbach04,Miyoshi11}. However, several studies reported non-detection of such periodicity \citep{Do09,Dexter14}, making the existence of any QPOs controversial. Subsequently, several-hour correlations have been detected in the submillimeter and NIR fluxes \citep{Dexter14,Meyer09,Witzel18}. \citet{Dexter14} has argued that the hourly timescale is comparable to the viscous timescales at the ISCO of Sgr A*.

Recent observations with the Atacama Large Millimeter/submillimeter Array (ALMA) have provided detailed information in regard to the spectral index \citep{Bower15,Bower19}, polarization \citep{Liu16a,Liu16b,Bower18}, and time variability \citep{Brinkerink15,Bower18}, owing to the high-sensitivity and stability of ALMA. In this Letter, we present and analyze high-quality light curves with a time resolution of $\sim 1$ min using ALMA at 230 GHz. The main objective of this study is confirming the short-timescale variability in the light curves. Since the short-timescale variability may be originated from the vicinity of the supermassive black hole, it will provide information as regards the property of Sgr A*.

\section{Observations and Data Reduction} \label{sec:obs}

\begin{deluxetable*}{clccccccccc}
\tablecaption{ALMA Band 6 Observations of Sgr A* Flux \label{tab:obs}}
\tablehead{
\colhead{Epoch} & \colhead{Date} & \colhead{Start Time\tablenotemark{a}} &
\colhead{$N_{\rm obs}$} & \colhead{Max $F_{\rm 217}$} & 
\colhead{Min $F_{\rm 217}$} & \colhead{Max $F_{\rm 219}$} & 
\colhead{Min $F_{\rm 219}$} & \colhead{Max $F_{\rm 234}$} & 
\colhead{Min $F_{\rm 234}$} &\\ 
\colhead{} & \colhead{(UT)} & \colhead{(UT)} & \colhead{} & \colhead{(Jy)} & 
\colhead{(Jy)} & \colhead{(Jy)} & \colhead{(Jy)} &
\colhead{(Jy)} & \colhead{(Jy)}
} 
\startdata
1 & 2017 Oct 5 & 23:40:42 & 44 & 2.958 & 2.749 & 2.954 & 2.748 & 2.969 & 2.757 \\
2 & 2017 Oct 7 & 00:16:15 & 45 & 3.276 & 3.042 & 3.275 & 3.046 & 3.327 & 3.096 \\
3 & 2017 Oct 8 & 23:52:35 & 44 & 3.174 & 2.901 & 3.176 & 2.897 & 3.228 & 2.932 \\
4 & 2017 Oct 10 & 00:06:17 & 45 & 3.317 & 3.131 & 3.317 & 3.129 & 3.370 & 3.186 \\
5 & 2017 Oct 11 & 00:10:15 & 45 & 3.340 & 3.150 & 3.337 & 3.143 & 3.391 & 3.164 \\
6 & 2017 Oct 11 & 22:50:01 & 44 & 3.014 & 2.853 & 3.016 & 2.860 & 3.044 & 2.882 \\
7 & 2017 Oct 14 & 22:45:59 & 44 & 3.017 & 2.896 & 3.012 & 2.902 & 3.048 & 2.935 \\
8 & 2017 Oct 17 & 00:03:21 & 45 & 2.811 & 2.461 & 2.821 & 2.455 & 2.828 & 2.467 \\
9 & 2017 Oct 18 & 00:35:31 & 44 & 2.225 & 2.004 & 2.231 & 1.999 & 2.206 & 1.977 \\
10 & 2017 Oct 19 & 23:56:49 & 45 & 3.053 & 2.903 & 3.068 & 2.909 & 3.098 & 2.918 \\
\enddata
\tablenotetext{a}{Start time of the on-source scan.}
\end{deluxetable*}

The Band 6 observations of the Sgr A region were carried out as a part of the ALMA Cycle 5 (2017.1.00503.S; PI: M.Tsuboi) over ten days between 2017 October 5 and 2017 October 20 (Table \ref{tab:obs}). The field of view included Sgr A* and the IRS 13E complex. The central frequencies of the four spectral windows (SPWs) were 217.5, 219.5, 234.0, and 231.9 GHz. The bandwidth and frequency resolution of the SPWs were 2.0 GHz and 15.625 MHz, respectively. Since the last SPW included the H30$\alpha$ recombination line, we used the other three SPWs for the continuum imaging of Sgr A*. J1924--2914 was observed as a flux and bandpass calibrator, while phase calibration was performed with J1744--3116 and J1752--2956. Detailed description of the analysis and the results of the IRS 13E complex from the same data have already been published by \citet{Tsuboi19}.

Data reduction was performed using the Common Astronomy Software Applications package (CASA). After performing the standard procedures for flux, gain, and bandpass calibrations, we applied the phase self-calibration method iteratively. To obtain light curves of Sgr A*, snapshot images of each SPW were produced with integration times in the range of 16 to 56 sec. These integration times were dependent upon the on-source observation time of each scan. The synthesized beam sizes with natural weighting were $\sim 0\farcs04 \times 0\farcs02$. Several on-source scans that were not interposed between two adjacent phase calibrator scans were flagged because calibrations might not have worked well for these data. We obtained 44 or 45 snapshot images ($N_{\rm obs}$ in Table \ref{tab:obs}) with a 70 min duration for each epoch. Flux densities and their uncertainties were measured by fitting them to the snapshot images with the CASA task ``imfit''. Cycle 5's technical capabilities assure that the accuracy of the absolute amplitude calibration is better than $10\%$ for the Band 6 observations, while the uncertainties of fitting are $\la 0.1\%$.

\section{Results} \label{sec:results}

\begin{figure*}[ht!]
\plotone{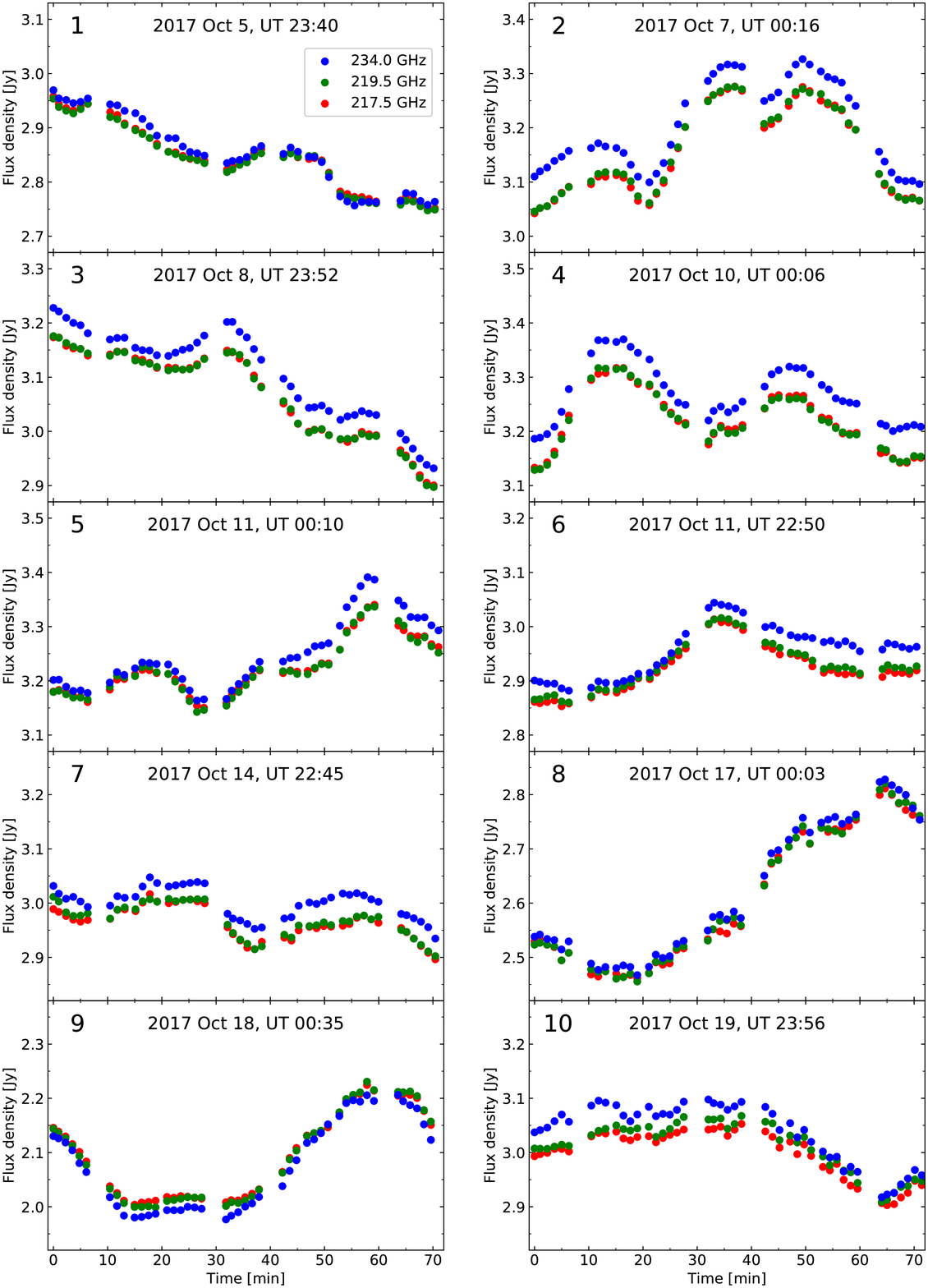}
\caption{Light curves of Sgr A* observed with ALMA at 217.5 (red), 219.5 (green), and 234.0 GHz (blue). The horizontal axis shows the time in minutes relative to the start time that is denoted on the top of each panel. The epoch number is shown on the top left corner. The errors are smaller than the symbol size. The absolute uncertainty in the flux scaling of each epoch is $\la 10\%$.  \label{fig:lightcurve}}
\end{figure*}

Figure \ref{fig:lightcurve} shows ten 70 min duration light curves of Sgr A* at 217.5, 219.5, and 234.0 GHz. All light curves exhibit clear intraday variations, and some appear to demonstrate periodicity. The flux density ranged from 2.0 to 3.4 Jy over 14 days, which is comparable to the submillimeter flux density value in the steady quiescent state \citep{Subroweit17}. The number of peaks in each light curve is 1--3, which corresponds to the variation timescale of a few tens of minutes. Within epochs 1 and 3, the short-timescale variations seem to be superposed on gradual flux changes, which is possibly due to variations with timescales longer than the duration time.

The average flux densities of 217.5, 219.5, and 234.0 GHz were $2.912\pm0.331$, $2.914\pm0.331$, and $2.939\pm0.348$ Jy, respectively. The largest amplitude of flux variation (14\%) can be seen during epoch 8, and ranges from $2.467\pm0.002$ to $2.828\pm0.003$ Jy at 234.0 GHz. The maximum and minimum flux densities across all of the observations were measured during epoch 5 and epoch 9, respectively. The observed maximum and minimum flux values of each epoch are summarized in Table \ref{tab:obs}.

The light curves are very similar between the three frequency bands. The majority of the flux densities at 234.0 GHz are higher than or equal to those at lower frequencies, resulting in positive spectral indices. However, for epoch 9 and part of epoch 1 the indices are negative. These results clearly indicate that the spectral index is also variable.

\section{Discussion}
\subsection{Variability Timescale}
\begin{figure}[t]
\plotone{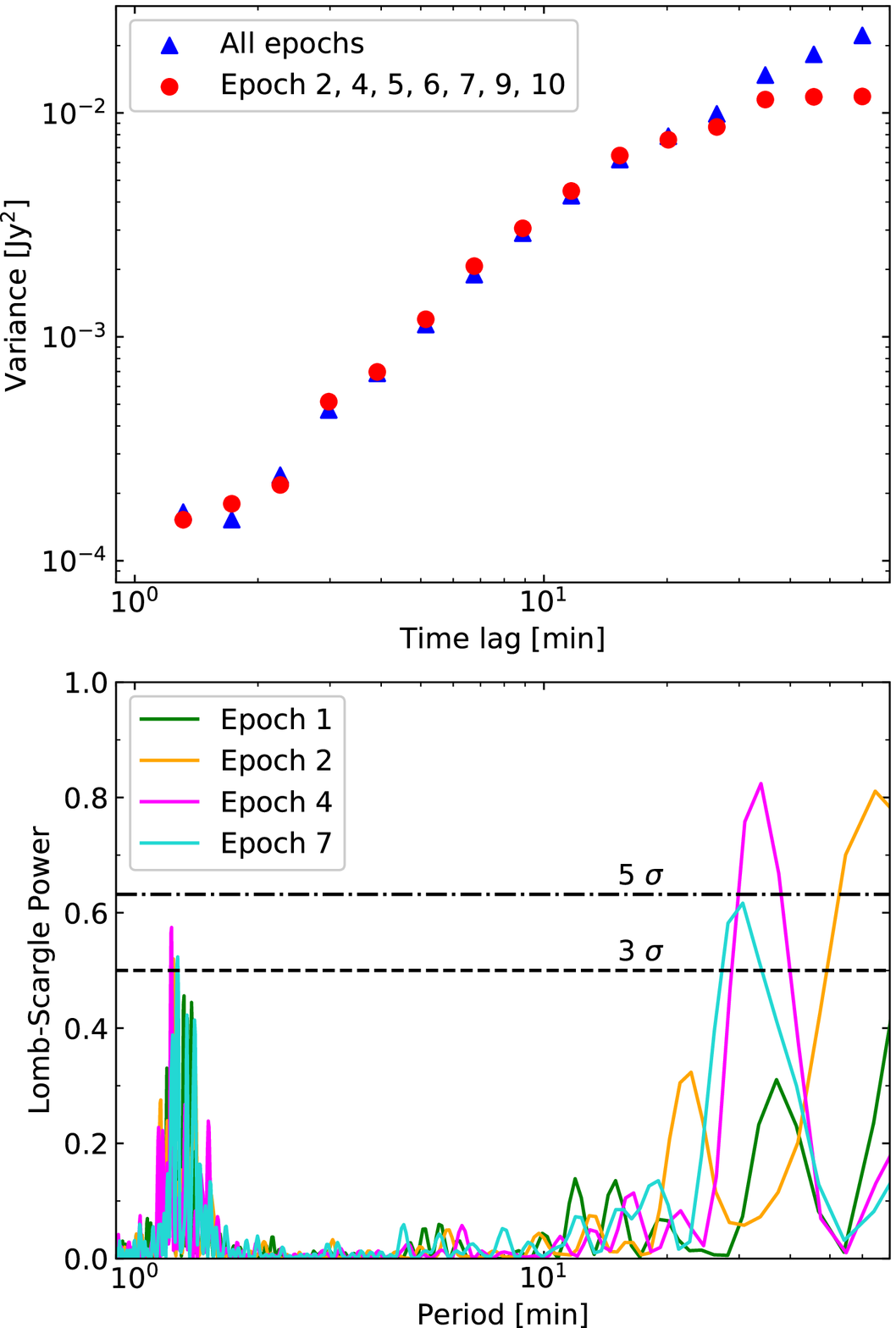}
\caption{(Top) Logarithmically binned structure functions of the light curves at 234.0 GHz. Red points show the structure function for the light curve concatenated from all epochs, and blue points depict it for all epochs except 1, 3 and 8. (Bottom) Lomb-Scargle periodogram for the 234.0 GHz light curves of epochs 1 (green), 2 (orange), 4 (magenta), and 7 (cyan). The false-alarm probabilities of $3\,\sigma$ and $5\,\sigma$ levels derived from the \citet{Baluev08} method are shown in the dotted and dashed lines, respectively.
\label{fig:structure_func}}
\end{figure}

To characterize the variability of Sgr A*'s flux in the 230 GHz band, we employed the first-order structure function, which is widely used to measure the characteristic timescale in light curves of active galactic nuclei and Sgr A* \citep{Simonetti85,Falcke99,Yusef-Zadeh11,Dexter14,Meyer09,Witzel18}. The first-order structure function $V(\tau)$ is defined as the variance:
\begin{equation}
V(\tau) = \langle [F(t+\tau) -F(t)]^2\rangle ,
\end{equation}
Where $F(t)$ is an arbitrary function of time $t$. The time lag $\tau$ at a maximum value of $V(\tau)$ corresponds to a characteristic timescale. We divided the variances into 15 bins equally spaced in logarithmic time lags up to 60 min. The structure functions derived from the concatenated light curve of all epochs at 234.0 GHz are shown in the top panel of Figure \ref{fig:structure_func} in blue. Each bin contains at least 48 variances. The structure functions of 217.5 and 219.5 GHz behave almost identically, although they are not shown in the figure.

The variance for the light curve of all epochs increases for time lags of up to 60 min. This indicates that variability of longer than 60 min exists in the light curves. Characteristic timescales of several hours were found in the submillimeter and NIR fluxes of Sgr A* \citep{Dexter14,Meyer09,Witzel18}. These hourly timescales are interpreted as being related to viscous timescales near the ISCO rather than orbital periods \citep{Dexter14}. The observed long-timescale variation may be of the same origin as the previously observed hourly scale variations.

Long-timescale variations are clearly seen in the light curves of epochs 1 and 3, and also possibly epoch 8. However, short-timescale variations have been superposed on long-timescale variations, being dominant in the other epochs. To identify the characteristic time of the short-timescale variation, we calculated the structure function from the light curves excluding epochs 1, 3 and 8. The result is also shown in the top panel of Figure \ref{fig:structure_func}. The variance increases up to the time lag of $\sim 20$ min, with almost the same slope of that derived from all epochs. The slope becomes flat for time lags larger than $\sim 30$ min. This changing point corresponds to the characteristic timescale. The variance asymptotically approaches the value of $\sim 10^{-2}$ Jy$^2$, indicating the amplitude of the short-timescale variations of $\sim 0.1$ Jy.

We also tried to derive the characteristic timescale by searching for periodicity in the light curves because structure functions may not be reliable for quantitative determination of timescales \citep{Emmanoulopoulos10}. To identify periodicity in the unevenly sampled light curves, we computed the Lomb-Scargle periodogram \citep{Lomb76,Scargle82,VanderPlas18}. The bottom panel of Figure \ref{fig:structure_func} shows the resultant periodograms for the light curves of epochs 1, 2, 4 and 7. The two periodograms of epochs 4 and 7, in which the light curves show sinusoidal shapes, exhibit prominent peaks at $\sim 30$ min. The heights of the peaks well exceed the $3\,\sigma$ level of the false-alarm probability \citep{Baluev08}. The epoch 1 periodogram shows a faint peak at $\sim 40$ min. The epoch 2 periodogram also has a faint peak at $\sim 20$ min and a significant peak at $\sim 60$ min. The periodograms from the other epochs have no significant peaks. Except for the 60 min peak in the epoch 2, each of these peaks is consistent with that expected from the number of intensity peaks in the corresponding light curve. We therefore conclude that the quasi-periodicity with timescales of a few tens of minutes up to an hour appears in the light curves. The periodogram analysis supports the results from the structure functions.
 
Here we consider the meaning of the few tens of minutes quasi-periodicity, assuming that it reflects a certain physical process. The orbital period at the ISCO of Sgr A* is $\sim 30$ min if the black hole does not rotate. Such orbital motions around Sgr A* have recently been detected in NIR flares with a period of $\sim 40$ min \citep{Gravity18b}. Therefore, our observed short-timescale variations possibly originate from the orbital motions near the ISCO of Sgr A*. In this case, its timescale varies depending on the orbital radius. The difference in the quasi-period between epochs may reflect different orbital radii of millimeter-wave emitting hotspots.

\subsection{Time Lag}
Previous studies have suggested that a flare observed at lower frequencies may be delayed from those at higher frequencies \citep{Yusef-Zadeh06,Yusef-Zadeh08}. To examine any time lag between the observed frequencies, we used the z-transformed discrete correlation function \citep[ZDCF;][]{Alexander97}. This method enabled us to estimate a cross-correlation function from unevenly sampled light curves. We obtained ten ZDCFs for the light curves at 217.5 and 234.0 GHz of each epoch. We estimated the time lags by fitting a quadratic function to the peaks of the ZDCFs.

All ZDCFs except for epochs 6 and 10 show absolute time lags of less than 1 min. The variation during epoch 6 at 234.0 GHz is delayed by $2.6\pm0.3$ min from that at 217.5 GHz. On the other hand, the variation within epoch 10 at 234.0 GHz precedes that at 217.5 GHz by $1.4\pm0.4$ min. These two lags seemingly do not depend on the flux densities or spectral indices. Considering the near-zero lags of the eight ZDCFs, we conclude that there is no coherent time lag between fluxes across the different observed frequencies.

The origin of time lags at lower frequencies is proposed to be due to the change in the optical depth of adiabatically expanding plasma blobs. We estimated an expected time lag between 234.0 and 217.5 GHz based on the model of \citet{vanderLaan66} and the following analyses of \citet{Yusef-Zadeh06} and \citet{Miyazaki13}. The expected time lag $\Delta T_{\rm 4-3}$ between $\nu_4 = 234.0$ GHz and $\nu_3 = 217.5$ GHz is calculated as

\begin{equation}
\Delta T_{\rm 4-3}  = \frac{\nu_4 ^A - \nu_3 ^A}{\nu_2 ^A - \nu_1 ^A}\times\Delta T_{\rm 2-1},
\end{equation}
where $A = -(p+4)/(4p+6)$ and $p$ is the index of the relativistic electron energy spectrum $[n(E) \propto E^{-p}]$ \citep{Yusef-Zadeh06}. Using the time lag $\Delta T_{\rm 2-1}$ between $\nu_2 = 43$ GHz and $\nu_1 = 22$ GHz of 30 min \citep{Yusef-Zadeh06,Yusef-Zadeh08} and $p=2$, $\Delta T_{\rm 4-3}$ becomes 1.4 min, meaning that the variation at 234.0 GHz precedes by 1.4 min from that at 217.5 GHz. This expected time lag is comparable to the time resolution of the ALMA observations and could be detectable.

One likely reason why we could not detect the time lag was that our observations were performed during the quiescent state of Sgr A*, and time delay is often detected in the flaring states. \citet{Miyazaki13} found no time lag between light curves at 90 and 102 GHz. They suggested the reason is that the blob is initially optically thin, or that the emission does not originate in expanding plasma but instead comes from orbiting hot plasma spots on the accretion disk. The observed orbital timescale variations are in favor of the orbiting hotspots model.

The time lag between different frequencies could also be explained in the context of a jet model \citep{Falcke09,Brinkerink15}. Following the analysis of \citet{Brinkerink15}, we can constrain the gas flow velocity $v_{\rm f}$ by combining the time lag with the source size difference between two observed frequencies.  Given the upper limit of the time lag of 1.5 min (an average of the time resolution of our observations), and the source size difference of 3.6 $\mu$as, we found $v_{\rm f} \ge 0.16c$. This is consistent with $v_{\rm f} = 0.77c$ that was derived from the data of VLA and ALMA \citep{Brinkerink15}. Therefore, the jet model can not be ruled out by our time lag analysis. Further simultaneous multiwavelength observations are necessary to examine the jet model.

\subsection{Spectral Index}
\begin{figure}[t]
\plotone{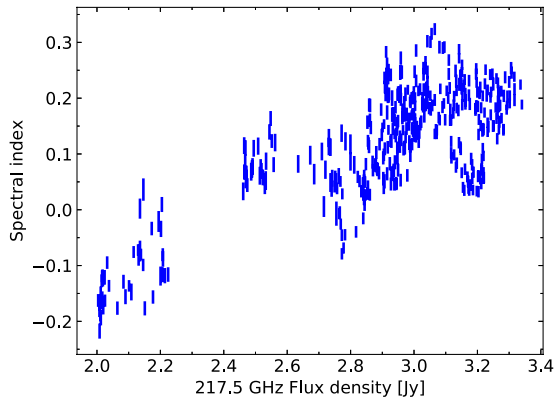}
\caption{Spectral index $\alpha$ (defined as $F_{\nu} \propto \nu^{\alpha}$) between 217.5 and 234.0 GHz as a function of the 217.5 GHz flux density. \label{fig:spectral_index}}
\end{figure}

We derived spectral indices between the flux densities at 217.5 and 234.0 GHz at all data points. The mean spectral index of all the data is $0.12\pm0.10$. The maximum and minimum values are $0.31\pm0.01$ and $-0.21\pm0.01$, respectively. In the NIR, several authors reported a variation in the spectral index dependent on the flux density \citep{Ghez05,Bremer11,Witzel18}. Figure \ref{fig:spectral_index} shows the spectral index dependence on the flux density at 217.5 GHz. A clear correlation that a higher flux density has a higher spectral index can be seen. 

In the radio band, some authors have already reported a trend that the spectral index becomes larger near the peak of the flux density in the centimeter \citep{Falcke99} and millimeter \citep{Miyazaki13} bands. In comparison with the spectral indices between 218.0 and 233.8 GHz derived from the ALMA Cycle 1 observations \citep{Bower15}, seven out of eight spectral indices agree with the trend of our result. 

The similar trend in the NIR suggests that the emissions in the 230 GHz and NIR bands originate from the same source. The dependence of the NIR spectral index on the flux density can be explained by the synchrotron emission in the optically thin frequency regime, with a variable exponential cutoff in the energy of the electron population due to synchrotron losses \citep{Eckart06,Bremer11,Witzel18}. This interpretation is, however, not applicable to the 230 GHz trend because a change in the cutoff frequency $\nu_0$ appears as a term of $\exp{[-(\nu/\nu_0)^{1/2}]}$, and $\nu_0$ of $\sim$ 10--100 THz does not alter spectral indices at the millimeter band coherently.

Thermal synchrotron emission with variable electron temperature may also explain the change of the spectral index in the submillimeter band by shifting the peak of the spectrum \citep{Bower15, Bower19,vonFellenberg18}. However, the thermal model solely cannot account for the NIR flux of Sgr A*, and an additional non-thermal component may contribute to it. Measuring variable SED must be necessary to unveil the emission mechanism of Sgr A*.

\section{Conclusions}
Using ALMA, we measured the flux densities of Sgr A* at 217.5, 219.5, and 234.0 GHz over ten days during the period from 2017 October 5 to 2017 October 20. We obtained high-quality light curves with a duration of 70 min. Two characteristic timescales were identified, one of a few tens of minutes and one of longer than 60 min. The long-timescale variation could correspond to the hourly scale ones that have been reported in the submillimeter and the NIR bands. The short-timescale variation likely originates from the orbital motions near the ISCO of Sgr A*. No time lag between 217.5 and 234.0 GHz was detected. We also confirmed the dependence of the spectral index on the flux density.

Further studies using archival data of ALMA will enable us to perform statistical analysis in the millimeter and submillimeter band. Comparing the properties in the millimeter and submillimeter flux variation with those in the NIR \citep[e.g.,][]{Witzel18} and X-ray fluxes \citep[e.g.,][]{Neilsen15} will be crucial to constrain the radiation models of Sgr A*. We expect that future ALMA observations with long duration times may detect a short-timescale variation with clear periodicity originating from the orbital motion close to Sgr A*. Gradual shortening of the period may indicate the accretion of emitting matter onto the supermassive black hole.
\newpage
\acknowledgments
This work was supported by Japan Society for the Promotion of Science (JSPS) Grant-in-Aid for JSPS Fellows Grant Number JP18J20450. Data analysis was carried out on the Multi-wavelength Data Analysis System operated by the Astronomy Data Center (ADC), National Astronomical Observatory of Japan (NAOJ). We thank T. Tsutsumi for the improvement of the calibrations, and the anonymous referee for helpful comments and suggestions. This paper makes use of the following ALMA data: ADS/JAO.ALMA\#2017.1.00503.S. ALMA is a partnership of ESO (representing its member states), NSF (USA) and NINS (Japan), together with NRC (Canada), MOST and ASIAA (Taiwan), and KASI (Republic of Korea), in cooperation with the Republic of Chile. The Joint ALMA Observatory is operated by ESO, AUI/NRAO and NAOJ. 

\vspace{5mm}
\facilities{ALMA.}

\software{CASA (version 5.5.0).}

\end{document}